\documentclass[review]{elsarticle}

\usepackage{lineno,hyperref,xcolor,amsmath}
\usepackage{framed,multirow}

\usepackage{amssymb}
\usepackage{amsmath}
\usepackage{latexsym}
\usepackage{subfig}
\usepackage{csquotes}

\usepackage{xcolor}
\definecolor{newcolor}{rgb}{.8,.349,.1}

\journal{Advances in Space Research}

\begin{document}

\begin{frontmatter}

\title{Monte Carlo Simulation of CRAND Protons Trapped at Low Earth Orbits}

\author{Ritabrata Sarkar\corref{cor1}}
\ead{ritabrata.s@gmail.com}
\cortext[cor1]{Corresponding author}

\author{Abhijit Roy}
\ead{aviatphysics@gmail.com}

\address{Indian Centre for Space Physics, 43 Chalantika, Garia Station Road,
Kolkata 700084, W.B., India}

\begin{abstract}
The Cosmic Ray Albedo Neutron Decay (CRAND) is believed to be the principal
mechanism for the formation of inner proton radiation belt -- at least for 
relatively higher energy particles. We implement this mechanism in a Monte
Carlo simulation procedure to calculate the trapped proton radiation at the low
Earth orbits, through event-by-event interaction of the cosmic ray particles in
the Earth's atmosphere and their transportation in the magnetosphere. We consider
the generation of protons from subsequent decay of the secondary neutrons from
the cosmic ray interaction in the atmosphere and their transport (and/or
trapping) in the geomagnetic field. We address the computational challenges
for this type of calculations and develop an optimized algorithm to minimize
the computation time. We consider a full 3D description of the Earth's
atmospheric and magnetic-field configurations using the latest available models. We
present the spatial and phase-space distribution of the trapped protons
considering the adiabatic invariants and other parameters at the low Earth
orbits. We compare the simulation results with the trapped proton flux
measurements made by {\it PAMELA} experiment at low Earth orbit and explain 
certain features observed by the measurement.
\end{abstract}


\end{frontmatter}


\section{Introduction} 
\label{sec:intro}
The study of the inner radiation belt is important, as this hostile radiation
environment impose crucial hazards to the satellites and space missions at low
and middle Earth orbits. While, the outer radiation belt is predominantly
composed of energetic electrons, the inner belt mainly consists of energetic
protons up to a few GeV, with a trace of e$^{\pm}$ and some other ions. The 
trapped protons in the radiation belt may have long residence time (up to 
several thousand years) and thus the accumulated radiation effect results from 
long time inputs from different sources like solar protons and Galactic Cosmic 
Rays (GCR). Several processes control the internal dynamics of the radiation 
belts. Radial diffusion provides the inward transport of the solar protons 
injected into the radiation belt at L $\gtrsim$ 2, during large solar proton 
events and magnetic storms \citep{Lore02, Huds04}. Losses of radiation belt 
particles are imposed by the processes like: ionization of the neutral 
atmosphere, energy loss to the plasmasphere and ionosphere, inelastic nuclear 
scattering and adiabatic energy and drift shell changes due to geomagnetic 
secular variations. For the radiation intensity at energies $\lesssim$ 100 MeV 
and for L $\gtrsim$ 1.3 are dominated by the solar protons \citep{sele07}. 
Otherwise, the protons from the $\beta$-decay of the free neutrons produced 
by the GCR interaction with the Earth's atmosphere, according to Cosmic Ray 
Albedo Neutron Decay (CRAND) mechanism \citep{sing58, farl71}, dominates the 
radiation belt.

There are some empirical trapped proton models available, to represent the inner
radiation belt, like: NASA AP8/AP9 model \citep{sawy76, gine13}, the model provided 
by the Institute of Nuclear Physics of Moscow State University \citep{gets91}, 
PSB97 model \citep{heyn99} etc. \citet{sele07} came up with a theoretical model 
of the inner proton radiation belt considering the physics and mechanism that 
populates the radiation belt. There are several satellite experiments that 
contributed to the understanding and modeling of the radiation belt like CRRES 
\citep{guss93}, SAMPEX/PET \citep{loop96}, TIROS/NOAA series \citep{hust96} etc. 
Lately, measurement of CR radiation at Low Earth Orbit (LEO) by {\it PAMELA} 
mission \citep{pico07}, provides important information on the high energy 
($\gtrsim$ 70 MeV) part of the trapped radiation at the lower region of the 
radiation belt.

In this work, we recreate the trapped proton distribution at LEO region,
considering realistic physical models of: Earth's atmosphere, magnetosphere,
GCR distribution, particle interactions in the atmosphere and their transport.
We embedded all the physical mechanisms in a single Monte Carlo (MC) simulation
framework. However, the total simulation procedure is segregated into different 
stages for convenience of the calculation and the outcome of each stage of the 
simulation flows into the following to get the final results. The details of the 
simulation procedure are discussed in Sec. \ref{sec:simu}. The spatial and 
spectral distribution of the secondary particles produced from the GCR 
interactions with the atmospheric nuclei depend on the rigidity cut-off of the 
GCR at different locations determined by the geomagnetic field distribution. 
The albedo neutrons produced in this way will have location dependency. The 
protons generated from the subsequent $\beta$-decay get trapped or escape 
according to their rigidity that depends on the energy and location of their 
generation. To include all these effects we perform the simulation starting 
right from the generation of GCR distribution at the Earth's orbit, their 
transportation into the Earth's atmosphere depending on their rigidity and 
subsequent interaction to produce secondary particles. Then, we consider the 
neutrons which escape into the space and decay to produce protons. The 
trajectories of these protons can be calculated individually from the equation 
of motion in the magnetic field. However, calculating the actual path due to 
Newton-Lorentz motion may lead to time consuming computation. So, instead, we 
adopted the guiding center equation to calculate the average track of the 
particles. In this work, we carry on with some simplified assumptions, 
like neglecting geomagnetic secular variation and considering an average 
heliospheric effect which, in principle, can limit some aspects of the 
calculation and will be discussed later at proper places (viz. Sec. 
\ref{ssec:norm}). We concentrate on the high energy part $\gtrsim$ 50 MeV 
to compare the simulation results with {\it PAMELA} measurement of trapped 
protons \citep{adri15}.

In the following Sec. \ref{sec:simu}, we outline the general simulation
procedure to describe the neutron generation from the GCR interaction with 
Earth's atmosphere, proton production from subsequent decay and their 
entrapment in the geomagnetic field. In Sec. \ref{sec:ana}, we present the 
analysis procedure and results from the calculations. We compare the simulation 
outcome with the recent measurements of the trapped protons by the {\it PAMELA} 
mission in Sec. \ref{sec:pamcom}. Finally we conclude the results from this 
work in Sec. \ref{sec:conc}.

\section{Simulation Procedure}
\label{sec:simu}
We considered an event-by-event simulation procedure using the MC technique to 
study the problem of trapped protons from the CRAND mechanism. The general 
scheme for the simulation is as follows. At first we calculate the neutron 
production in the atmosphere due to the interaction of the CR particles. These 
free neutrons, therefore decay to produce protons according to the 
$\beta$-decay process. Finally, the protons either get trapped into the 
radiation belt, transported off to the outer space or absorbed at the Earth -- 
according to their position of generation, rigidity, the direction of motion 
(pitch angle) and the geomagnetic field distribution. In the following 
subsections we briefly discuss these steps of the simulation procedure.

\subsection{Neutron production in the atmosphere}
\label{ssec:nprod}
To calculate the neutron production from the CR interaction in the atmosphere,
we considered a detailed MC simulation using Geant4 simulation toolkit
\citep{agos03}. The details of this simulation procedure can be found in
\citet{sark20}. However, for the sake of completeness we briefly mention the 
key aspects of the simulation here. We implemented a full 3D model of the
atmosphere and magnetosphere in Geant4 simulation environment, with proper
distribution of primary GCR particles at the Earth's orbit and using a list of 
interaction processes suitable for the energies and phenomena considered 
in the simulation.

The atmospheric matter distribution which depends on the location, time and 
other conditions like solar activity, is realized using the NRLMSISE-00
standard atmospheric model \citep{pico02}. The atmospheric distribution is
limited up to 100 km from the Earth's surface since most of the CR interactions 
effectively take place below this level to produce neutrons, which is our prime 
concern for this step of the simulation. The whole atmosphere is subdivided 
into 100 concentric spherical layers with the thickness equal in logarithmic scale 
of altitude to implement the gradual material distribution. However,
with an approximation, to evade the complexity in the atmospheric geometry 
construction, we did not implement the zonal and meridional distribution of 
the atmospheric constituents in the current simulation. We neither considered 
the temporal variation of the model parameters. Instead, we considered global 
atmospheric parameters at each layer corresponding to the tropical region 
(23$^{\circ}$N, 88$^{\circ}$E) and at a fixed time in the month of May in 2016 
\citep{sark20}.

The magnetic field distribution is also embedded in the simulation environment
along with the atmospheric matter distribution which directly takes care of the
rigidity cut-off of the charged primary particles and defines the trajectories
of the primary and secondary charged particles. The inner magnetic field due to
the Earth's inherent magnetism extends up to about 4 Earth radii ($R_E$) from 
the surface and is represented by the 12th generation IGRF model \citep{theb15} 
with proper input parameters at moderate solar condition (during 2016 at 
declining phase of 24$^{\rm th}$ solar cycle; vide \citet{sark20}). The 
external magnetic field depends on the inter-planetary magnetic field and 
solar conditions, is considered up to 25 $R_E$ defined by using the Tsyganenko 
model \citep{tsyg16}. Here also, in defining the magnetic field
model, we did not consider the temporal variation due to solar activity or the 
secular variation of the geomagnetic field, instead used the average field 
distribution corresponding to the above mentioned time.

For the primary GCR flux distribution, we considered only the most abundant
components, i.e., protons (H) and helium nuclei (He) which constitute about
99\% of the GCR flux. We used up-to-date spectral information of the primary CR
particle flux using the model for local interstellar spectrum given by
\citet{vos15}. This model is conceived by using the {\it Voyager I} data of the 
post-heliopause CR observation complimented by {\it PAMELA} and {\it AMS02}
data. The details of the primary CR flux distribution is given in
\citet{sark20}. To maintain the efficiency of the simulation, we generated an
isotropic distribution of the charged particles from a geocentric spherical
surface at 500 km above the Earth's surface and backtracked them to reach an 
outer surface at 25 $R_E$ in presence of the magnetic field distribution. We 
select only those particles that reach the outer sphere as the allowed CR 
particle tracks at 500 km and proceed with the simulation for atmospheric 
interactions. Thus, we consider the modulation of the primary CR at the Earth's 
orbit by the geomagnetic field surrounding Earth till 25 $R_E$. We restricted 
the primary generation in the energy range of 0.1--800.0 GeV/n. We simulated 
3.5 $\times$ 10$^5$ particles (both for H and He) that survived the rigidity 
cut-off at 500 km above the Earth's surface in the back-tracking method, to 
interact with the atmosphere. To achieve this number of particles to interact 
in the atmosphere, the simulation discarded about 10$^6$ H and 1.5 $\times$ 
10$^6$ He particles. Simulation of these number of events was required to 
validate simulation results with the {\it AMS02} observations with comparable 
significance \citep{sark20}.

To simulate the interaction of high energy particles with the atmosphere we used 
the reference physics list (QGSP physics list with Binary Cascade model {\it
QGSP\_BIC\_HP}) provided in Geant4, which optimally covers the interaction
processes relevant here.

\begin{figure}
  \centering
  \noindent{\includegraphics[width=0.8\textwidth]{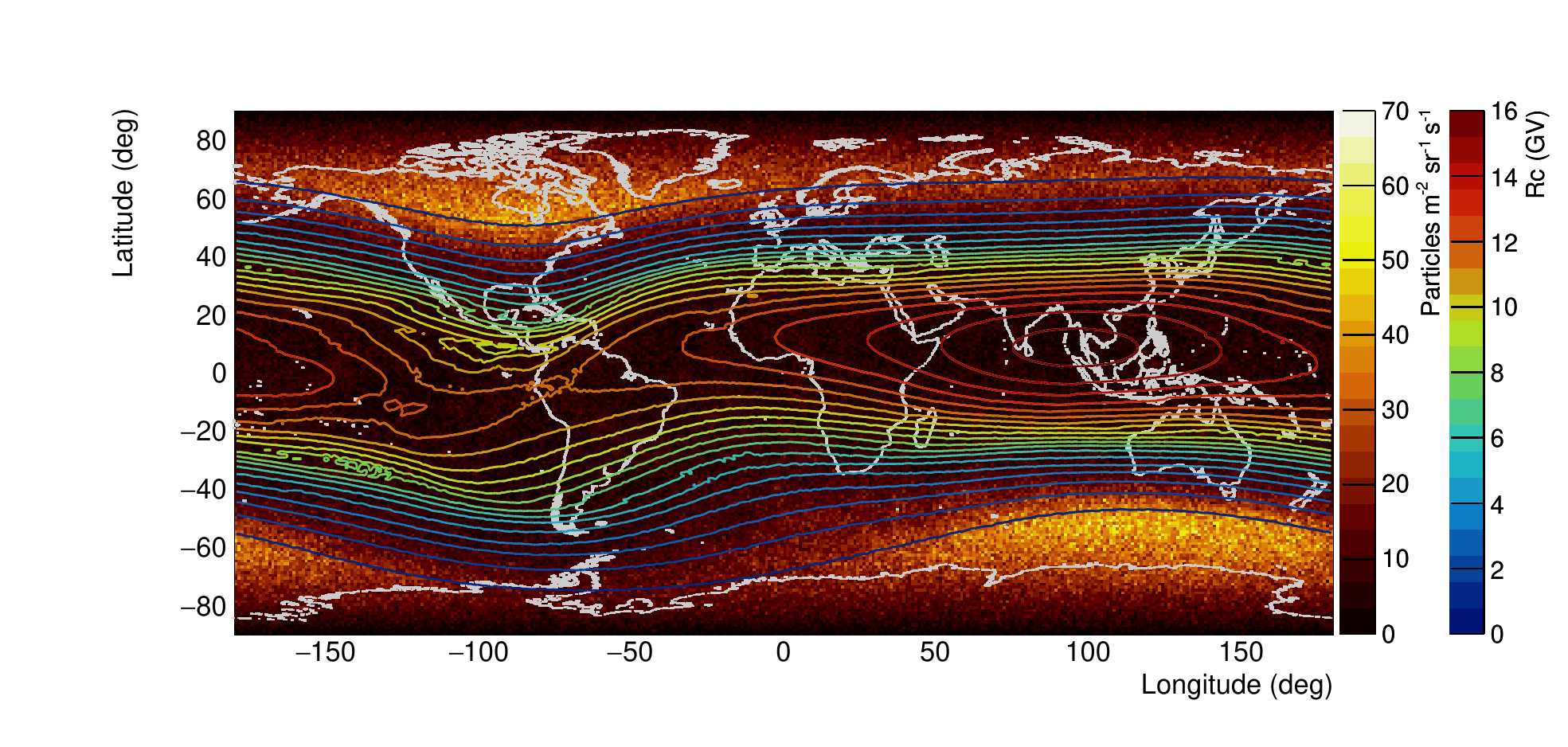}}
  \caption{Positional distribution of the secondary neutron flux at the top of
  the atmosphere (100 km from Earth's surface) as obtained from the simulation,
  due to primary cosmic rays (H and He). The contour plot of vertical rigidity cut-off
  values at different locations along with the continental coast line map for
  physical location guidance are also superimposed in the plot.}
   \label{fig:neutpos}
\end{figure}

Here, we are particularly interested in the neutrons produced due to CR
interactions in the atmosphere, which move outwards from the atmosphere. So, we
calculated the secondary neutron distribution at the spherical surface at 100 km
above the Earth's surface, which are outgoing from the atmosphere into the space 
(albedo neutrons). The spatial distribution of the secondary neutrons depend on 
the rigidity cut-off of the primary CR particles and hence on the geomagnetic 
field distribution. The spatial distribution (in geographic latitude
($\theta$)/longitude ($\phi$)) of the neutrons at 100 km above the Earth's 
surface is shown in Fig. \ref{fig:neutpos} along with vertical rigidity cut-off 
values at different locations. The apparent drop-off in the 
neutron flux toward the poles are due to the coordinate ($\theta$-$\phi$) we 
chose for the plotting. (For a uniform positional distribution on a spherical 
surface $\theta$ is not isotropic in [0, $\pi$] rather $\cos\theta$ is uniform
in [0, 1].) The overlaid contour plot is also calculated 
according to the chosen coordinates and this note on the distortion of the 
plots in $\theta$-$\phi$ coordinate is also valid for other subsequent 
plots (viz. Fig. \ref{fig:spdist} and \ref{fig:spdistpam}). The positional 
distribution, as well as the directional and 
spectral distributions of the secondary neutrons are used to generate the CRAND 
protons described in the next Sec. \ref{ssec:ndecay}. We used the calculated 
positional distribution shown in Fig. \ref{fig:neutpos} to generate random 
neutron positions in $\theta$/$\phi$ for the next step of the simulation.

\begin{figure}
  \centering
  \noindent\subfloat[]{\includegraphics[width=0.49\textwidth]{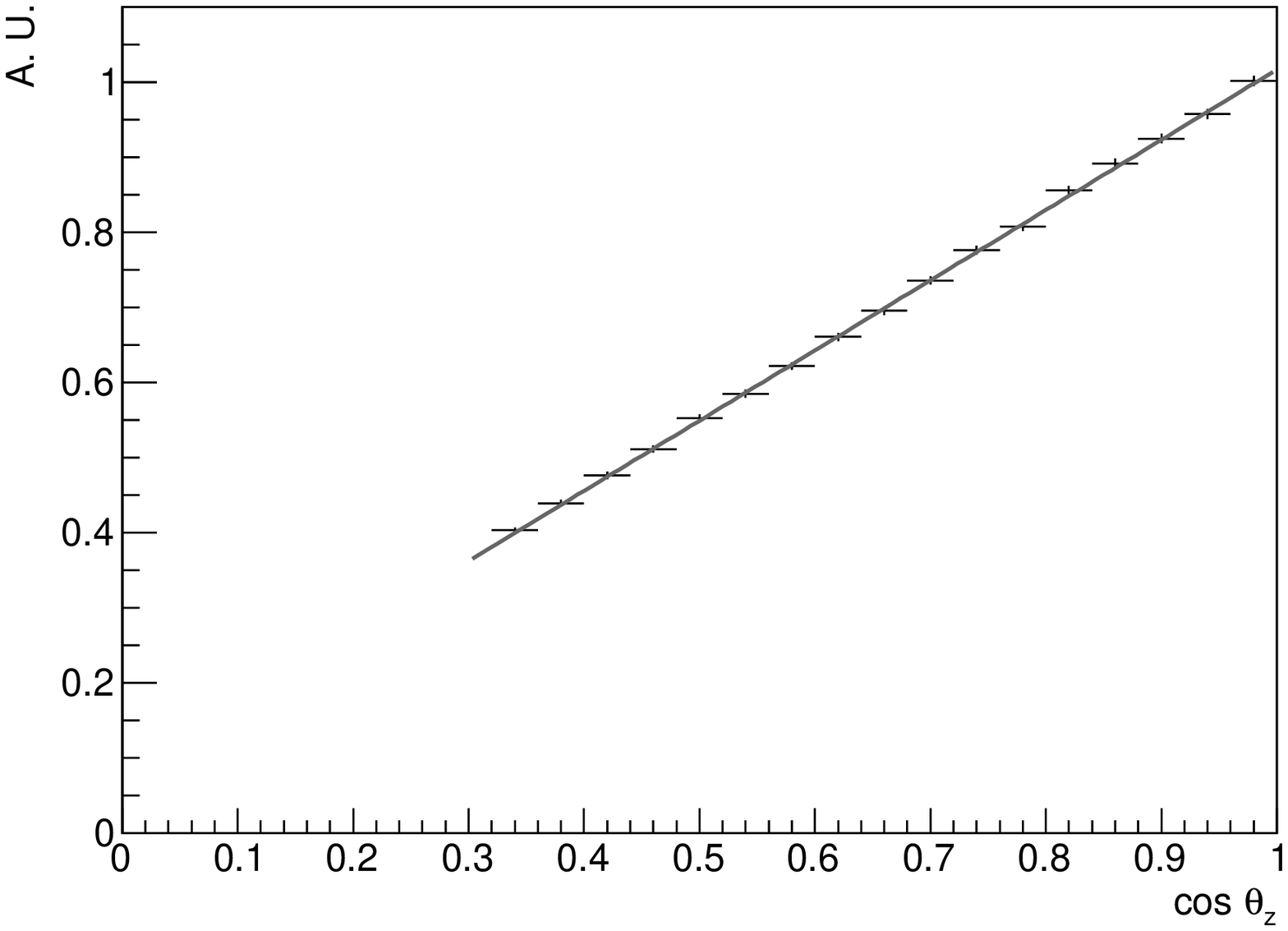}}
  \noindent\subfloat[]{\includegraphics[width=0.49\textwidth]{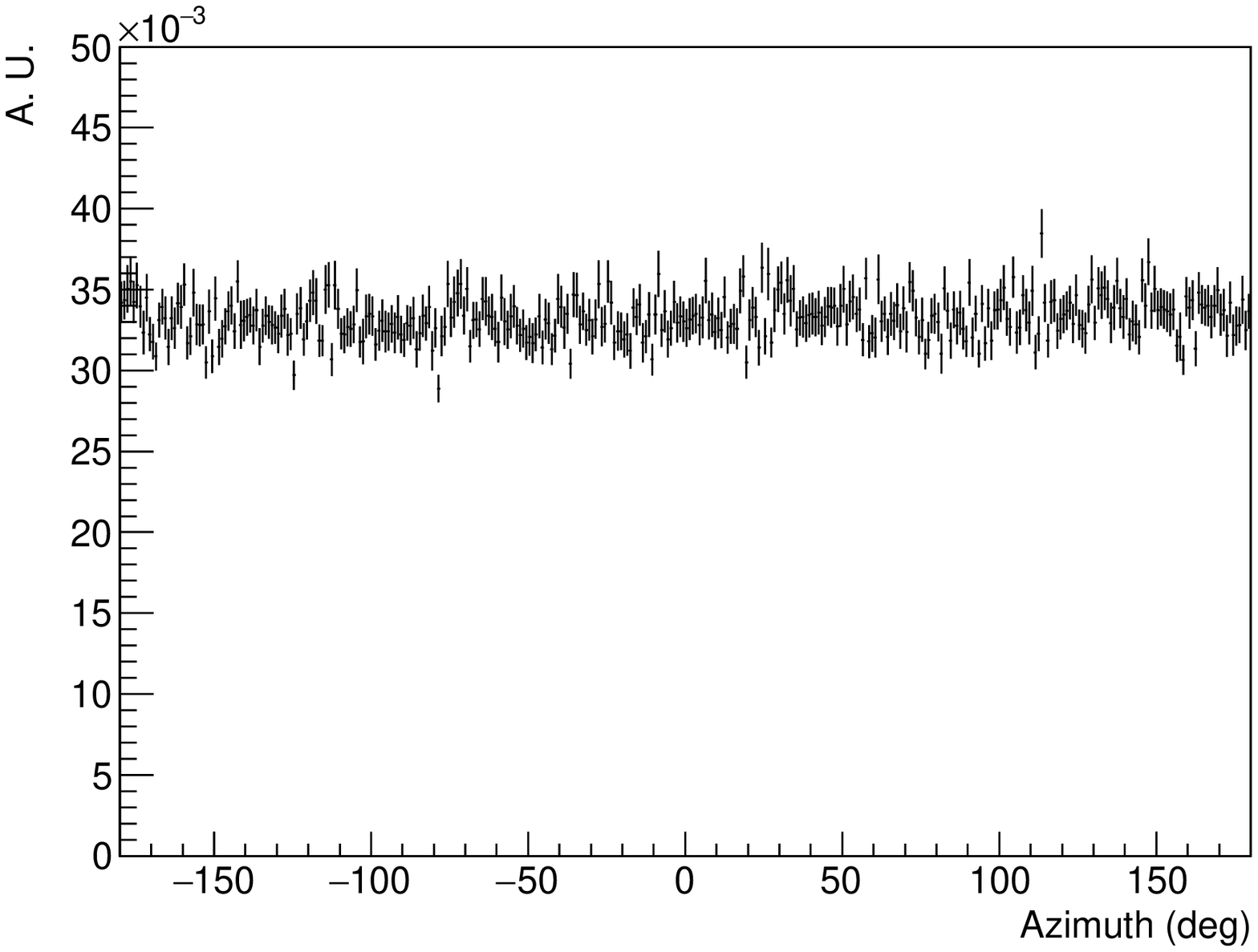}}
  \caption{(a) Neutron directional distribution as a function of $\cos\theta_Z$. 
  (b) Neutron directional distribution as a function of the azimuthal angle 
  ($\phi_Z$).}
  \label{fig:neutdir}
\end{figure}

The propagation (or momentum) direction of the produced neutrons are also
important in this concern; since the pitch angle of the protons, generated from
the decay of these neutrons, depends on their direction. We have shown this 
directional dependence in terms of $\theta_Z/\phi_Z$ angle in Fig. 
\ref{fig:neutdir}. Where $\theta_Z$ is the angle between propagation direction 
and zenith direction at the position of the neutron and $\phi_Z$ is the 
azimuthal angle about the zenith direction and with respect to local north. The 
cut-off that appeared in the $\cos\theta_Z$ distribution of the albedo neutron is 
due to the fact that, we generated the primary H and He from a spherical 
surface at 500 km above the Earth's surface while the extent of the atmosphere 
is only up to 100 km. So, only the primaries inside a limited solid angle 
(about the zenith direction) among all those isotropically generated from the 
spherical surface, can interact with the atmosphere. While this particular 
configuration of the primary generation was originally considered to serve some 
general purpose to calculate space and atmospheric radiation due to cosmic rays 
as discussed in \citet{sark20}, this may have some minor effect on the low 
pitch angle trapped particle population as will be discussed later at proper 
place (viz. Sec. \ref{ssec:tracvar} and \ref{ssec:adinv}).

\begin{figure}
  \centering
  \noindent\includegraphics[width=0.6\textwidth]{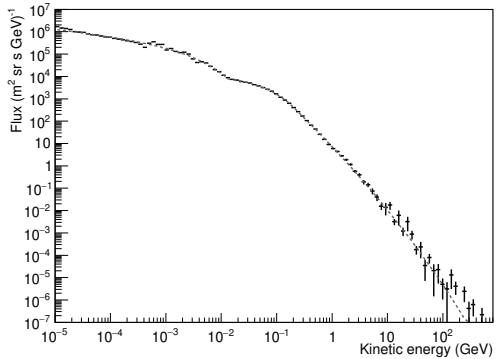}
  \caption{Albedo neutron flux distribution at the top of the atmosphere (at 100
  km) due to cosmic ray (H and He) interactions in the atmosphere. The spectrum is
  fitted (dashed line) with an empirical function given by Eq. \ref{eq:neutspec}.}
   \label{fig:neuteng}
\end{figure}

The energy distribution of the produced neutron flux is shown in Fig.
\ref{fig:neuteng}. This spectrum is fitted with an empirical function with
three modified cut-off power-laws (at three different energy intervals) of the
form:
\begin{equation}
  \frac{dN}{dE} = \sum\limits_{i=1}^3 A_i
  E_i^{-\alpha_i}exp\left(-\left(\frac{E_i}{B_i}\right)^{\beta_i}\right),
  \label{eq:neutspec}
\end{equation}
where the energy $E_i$ is in the ranges
$(\leftarrow:1.5\times10^{-2}:2.0\times10^{-2}:\rightarrow)$ GeV. The three sets
of ($A_i$, $\alpha_i$, $B_i$, $\beta_i$) values obtained from the fit are 
{(4.06712 $\times$ 10$^{+4}$, 3.03 $\times$ 10$^{-1}$, 2.27 $\times$ 10$^{-3}$, 
6.14 $\times$ 10$^{-1}$); (7.77445 $\times$ 10$^{+4}$, -3.07 $\times$ 10$^{-1}$, 
1.30 $\times$ 10$^{-2}$, 5.67 $\times$ 10$^{-1}$); (1.45930 $\times$ 10$^{+3}$, 
1.75, 4.84 $\times$ 10$^{-5}$, 1.68 $\times$ 10$^{-1}$)} respectively for three 
energy ranges. We used this function to generate the neutrons for subsequent 
generation of CRAND protons described in the next Sec. \ref{ssec:ndecay}.

It might be worth to mention here that actually the energy and 
directional distribution of the secondary neutrons may depend on the location 
(with geomagnetic latitude through rigidity cut-off and also to some extent due 
to atmospheric distribution). So, to achieve a more realistic neutron 
generation model in this phase of the simulation, one might incorporate the 
location dependent energy and directional distribution. In the current 
simulation, for the sake of simplicity, we adopted the averaged distribution 
in energy and direction over the whole area. However, we confirmed that at 
different geomagnetic latitude the shape of the energy spectra remains more or 
less identical. The variation in amplitude of the omnidirectional energy 
spectrum is taken care by the location dependent particle distribution shown 
in Fig. \ref{fig:neutpos}. We also observed only a nominal deviation in 
$\theta_Z$ distribution at the equatorial region (0--30$^{\circ}$ geomagnetic 
latitude) with respect to the average plot shown in Fig. \ref{fig:neutdir}a. 
We estimated about 20\% lower relative flux at the higher $\theta_Z$ (or lower 
$\cos\theta_Z$, i.e., away from zenith) end in this comparison.

\subsection{Protons from albedo neutron decay}
\label{ssec:ndecay}
To generate the protons from the CRAND source, we used the spatial, directional
and spectral distribution of albedo neutrons shown in Fig. \ref{fig:neutpos},
\ref{fig:neutdir} and \ref{fig:neuteng}. At first we generated random neutrons
according to these distributions on a spherical surface at 100 km above Earth's
surface. Then protons were generated according to free neutron $\beta$-decay
scheme. We randomly sampled the neutron decay time ($t_D$) for each neutron from
the decay time distribution function:
$\frac{t_D}{\tau_n}\, exp\left(\frac{-t_D}{\tau_n}\right)$, 
where, $\tau_n$ = 881.5 s is the neutron mean life time. Then we calculated the
position of generation of the proton from the distance travelled by the neutron
before its decay as: $v_n\, \gamma\, t_D$, where, $v_n$ is neutron velocity
(calculated from the kinetic energy and track direction) and $\gamma$ is the
relativistic time dilation factor. For the sake of simplicity, we preserved the
direction and kinetic energy of the protons same as the neutrons. In the 
original simulation calculation we considered all the protons generated inside 
the magneto-pause (10 $R_E$) boundary. However, considering our current purpose
of calculating trapped protons and comparing with {\it PAMELA} measurement 
\citep{adri15}, we enforced two conditions limited by the measurement -- take 
account of only those protons which: (i) have kinetic energies in the range of 
50 MeV to 4 GeV and (ii) are generated below 2 $R_E$. The first limit is 
obvious due to the energy threshold of the instrument and maximum energy 
of the trapped particles found by the measurement. The second limit imposed 
here considering the fact that {\it PAMELA} can observe the stably trapped 
protons only for L-shell values in between $\sim$1.18 $R_E$ up to $\sim$1.7 $R_E$ 
\citep{adri15}.

\subsection{Proton transportation in the geomagnetic field}
\label{ssec:ptrans}
The trajectory of the protons produced from the decay of neutrons are guided by the
geomagnetic field according to their momentum and position. These particles will
then be trapped in the magnetic field or deflected (into the outer space
or towards Earth) depending on the mentioned parameters and their tracks can be
traced using numerical calculations \citep{smar00, smar05}.

However, the calculation of the actual charged particle tracks due to its 
Newton-Lorentz motion in the magnetic field can be computationally expensive.
So, instead, we considered the guiding center motion of the particles, while the
directional information of the original track is preserved by the pitch angle
information. An offhand calculation (with 1000 proton tracks of 50 MeV energy)
shows a reduction of the tracking time to $\sim$ 37\%.

The equations of motion of the guiding center are given by \citep{oztu12}:
\begin{equation}
  \frac{d \mathbf R}{dt} = \frac{\gamma m v^2}{2 q B^2}
  \left(1+\frac{{v_{\parallel}}^2}{v^2}\right) \hat{\mathbf b} \times {\mathbf
  \nabla} B + v_{\parallel} \hat {\mathbf b},
  \label{eqn:gce1}
\end{equation}

\begin{equation}
  \frac{d v_{\parallel}}{dt} = -\frac{\mu}{\gamma^2 m} \hat{\mathbf b} \cdot
  {\mathbf \nabla} B.
  \label{eqn:gce2}
\end{equation}
Here $\mathbf R$ is the guiding center position of the particle track, which
moves in the local magnetic field with magnitude $B$ and directional unit vector
$\hat{\mathbf{b}}$. $m$, $q$, $\gamma$, $v$ and $v_{\parallel}$ are the particle
(proton) mass, charge, Lorentz factor, speed and component of velocity parallel
to the magnetic field direction respectively. $\mu = \frac{\gamma^2 m
{v_{\perp}}^2}{2 B}$ (with $v_{\perp}$ the perpendicular component of the
particle velocity), is the magnetic moment of the current generated by the
circular motion of the charged particle in the magnetic field.

We calculated the guiding center trajectories of the particle motion in the
geomagnetic field by integrating the Equations \ref{eqn:gce1} and
\ref{eqn:gce2}. We considered the adaptive integration technique using
{\it boost::numeric::odeint} library \citep{boost} with {\it Cash-Karp}
algorithm ({\it runge\_kutta\_cash\_karp54}) for stepper. The initial values
supplied to the integration method are: the position of the generated proton
from the neutron decay, the velocity component of the particle parallel to the
magnetic field and initial stepping time depending on the particle velocity to
travel 500 km distance along the guiding center.

However, we record only the stably-trapped particle trajectories satisfying the
conditions that, while drifting through the full rotation around the Earth
($360^{\circ}$ change of azimuthal angle of the trajectory; we are using
geocentric coordinate system) it does not reach: (i) within 40 km from Earth's
surface (absorbing atmospheric limit) and (ii) the magnetopause limit at 
10 $R_E$. Both of these two conditions corresponds to the quasi-trapped or
un-trapped particles and are not considered in the present work.

The termination of the stably-trapped trajectory integration depends on the
morphology of the drift trajectory. Keeping in mind, the latitude/longitude
grid resolution of the order of a degree, during the spatial analysis of the
trajectory data, we intend to keep an azimuthal distance of $\sim 1^{\circ}$ 
between two successive crossing points of the trajectory with the latitude 
(say at magnetic-equator). Then, we calculate the number of full drifted 
rotation of the trajectory required to cross the equator at least 360 times. 
So the trajectories with higher drift speed (with larger separation between 
successive crossings) are traced for longer time. This is done for better
interpolation of the particle location on the drift shell. The overall 
information of each trajectory is saved for the analysis purpose described 
in the next section.

We simulated a total of 10$^6$ events in which protons generated from the
neutron decay are trapped in the geomagnetic field, satisfying the aforementioned
trapping conditions. To get this number of trapped particles, the simulation
discarded another 2793176 events which did not satisfied the the trapping 
conditions and correspond to quasi-trapped or untrapped particles and not been 
considered further.

\section{Data Analysis and Results}
\label{sec:ana}

\subsection{Physical parameters}
\label{ssec:tracvar}
\begin{figure}
  \centering
  \noindent\subfloat[]{\includegraphics[width=0.49\textwidth]{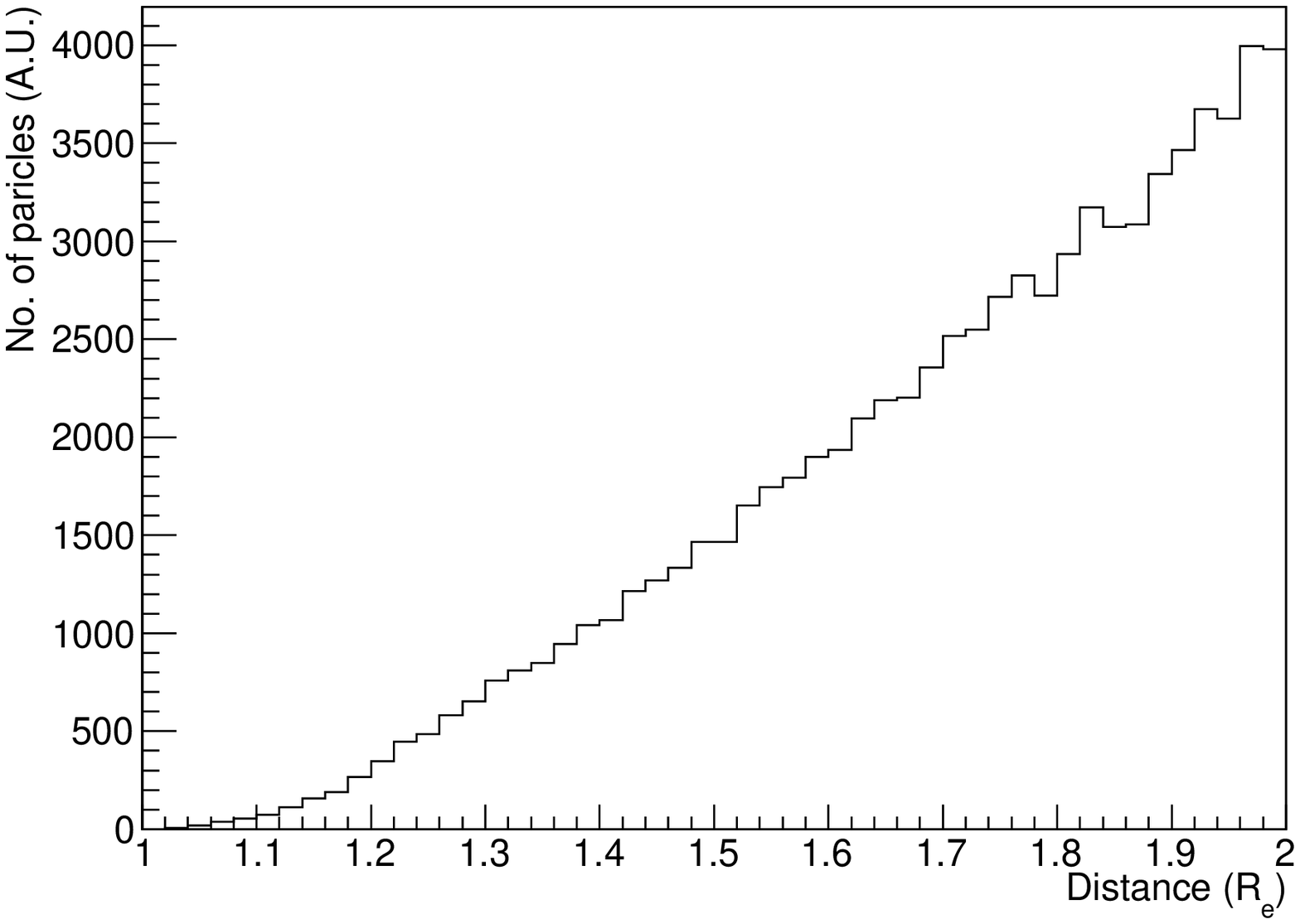}}
  \noindent\subfloat[]{\includegraphics[width=0.49\textwidth]{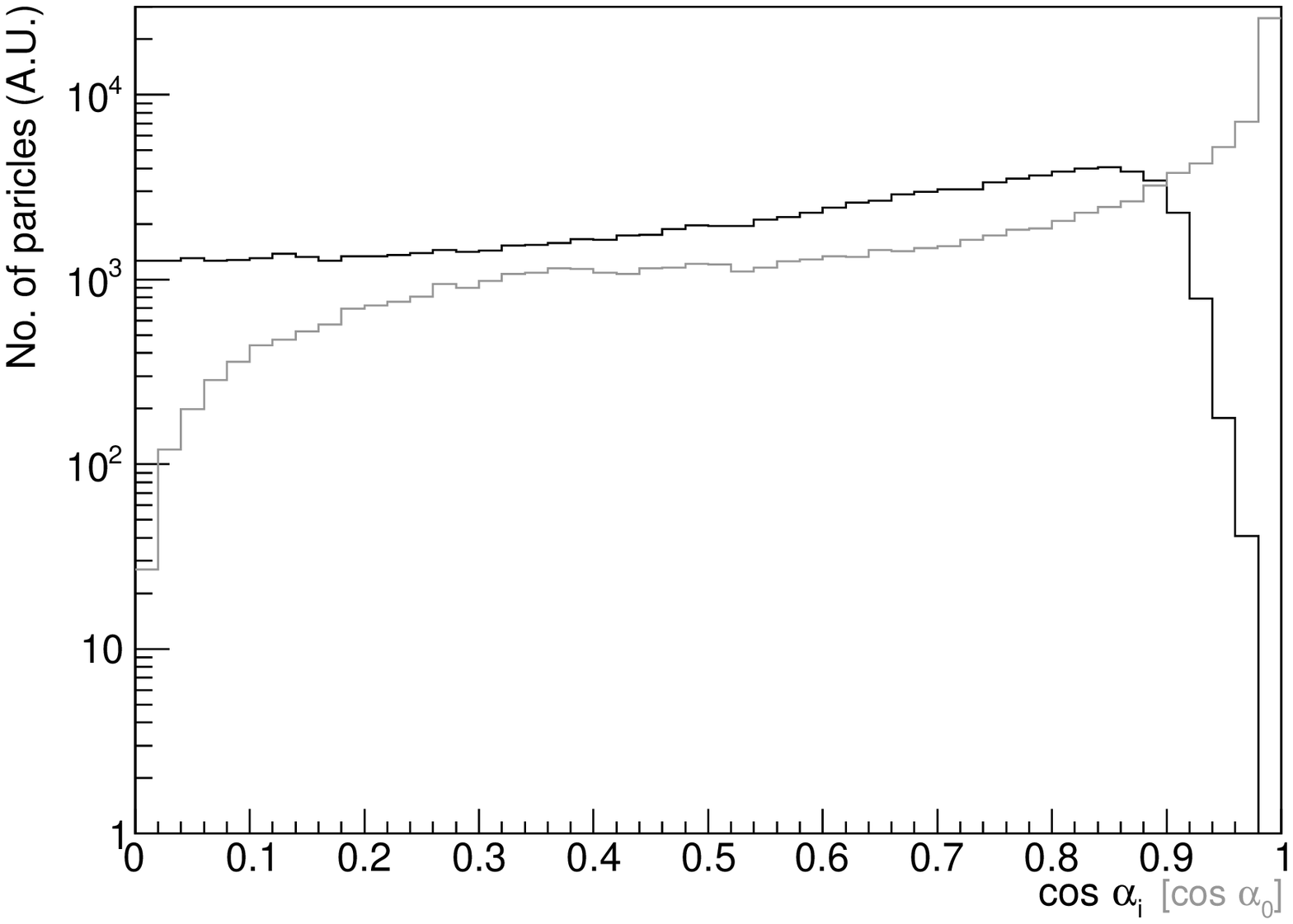}}
  \caption{(a) Trapped proton distribution as a function of the distance of initial 
  position of the trajectories. (b) Trapped proton distribution as the function of 
  cosine of the initial pitch angle (black) and the equatorial pitch angle (gray).}
  \label{fig:padist}
\end{figure}

At first, we check the three most important parameters at the generation point of
the proton trajectories: position of generation, direction of propagation and 
kinetic energy of the particle --- to look into the facts of delay in neutron 
decay, coupling of the proton with the geomagnetic field and energetics of the 
particle distribution. The initial position distribution of the trapped protons 
is shown in Fig. \ref{fig:padist}a, which shows the distance of proton 
generation from the Earth's center (in this simulation we considered this 
distance only up to 2 $R_E$). It is evident from this figure that the number 
of CRAND generated protons are increasing beyond 2 $R_E$, but, as we have 
already mentioned earlier, this limit is imposed in the prospect of comparing the
results with {\it PAMELA} measurement. We show the overall distribution of the 
initial pitch angle ($\alpha_i$; i.e., at the position of the proton 
generation) and the equatorial pitch angle ($\alpha_0$) of the trapped protons 
in Fig. \ref{fig:padist}b. The $\alpha_0$ values are calculated directly from the 
magnetic field information of the trajectory at the mirror points ($B_m$) and 
magnetic equator ($B_0$) as $\alpha_0$ = $\sin^{-1}\sqrt{\frac{B_0}{B_m}}$, 
averaged over the whole drift trajectory. The higher population near 
lower pitch angle (i.e., higher probability of particles having velocity 
nearly along or opposite to the magnetic field lines) can be explained from the 
following facts. The neutrons are generated in majority near the polar regions 
(where the inclination of the field lines are larger), as evident from Fig. 
\ref{fig:neutpos}. In addition, more particles are directed towards the zenith 
as suggested by the distribution of the momentum directions (Fig. 
\ref{fig:neutdir}). However, the scarcity of the protons with extremely low 
values of initial pitch angle (parallel or anti-parallel to the magnetic field) 
is due to the effects both from position distribution of the neutron generation 
and simulation artifact arose from primary particle generation surface and 
atmospheric distribution consideration as mentioned in Sec. \ref{ssec:nprod}.

\begin{figure}
  \centering
  \noindent\subfloat[]{\includegraphics[width=0.49\textwidth]{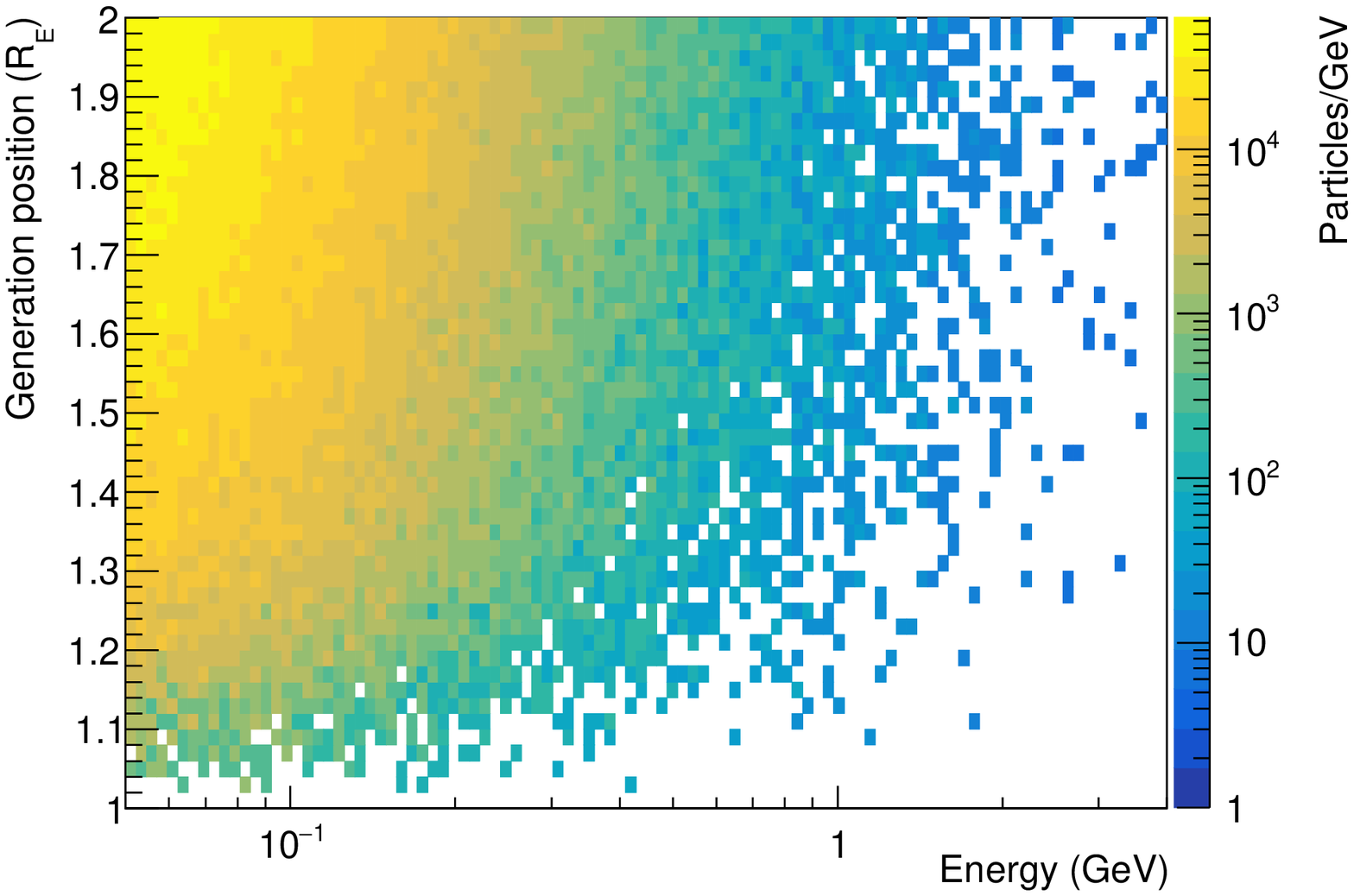}}
  \noindent\subfloat[]{\includegraphics[width=0.49\textwidth]{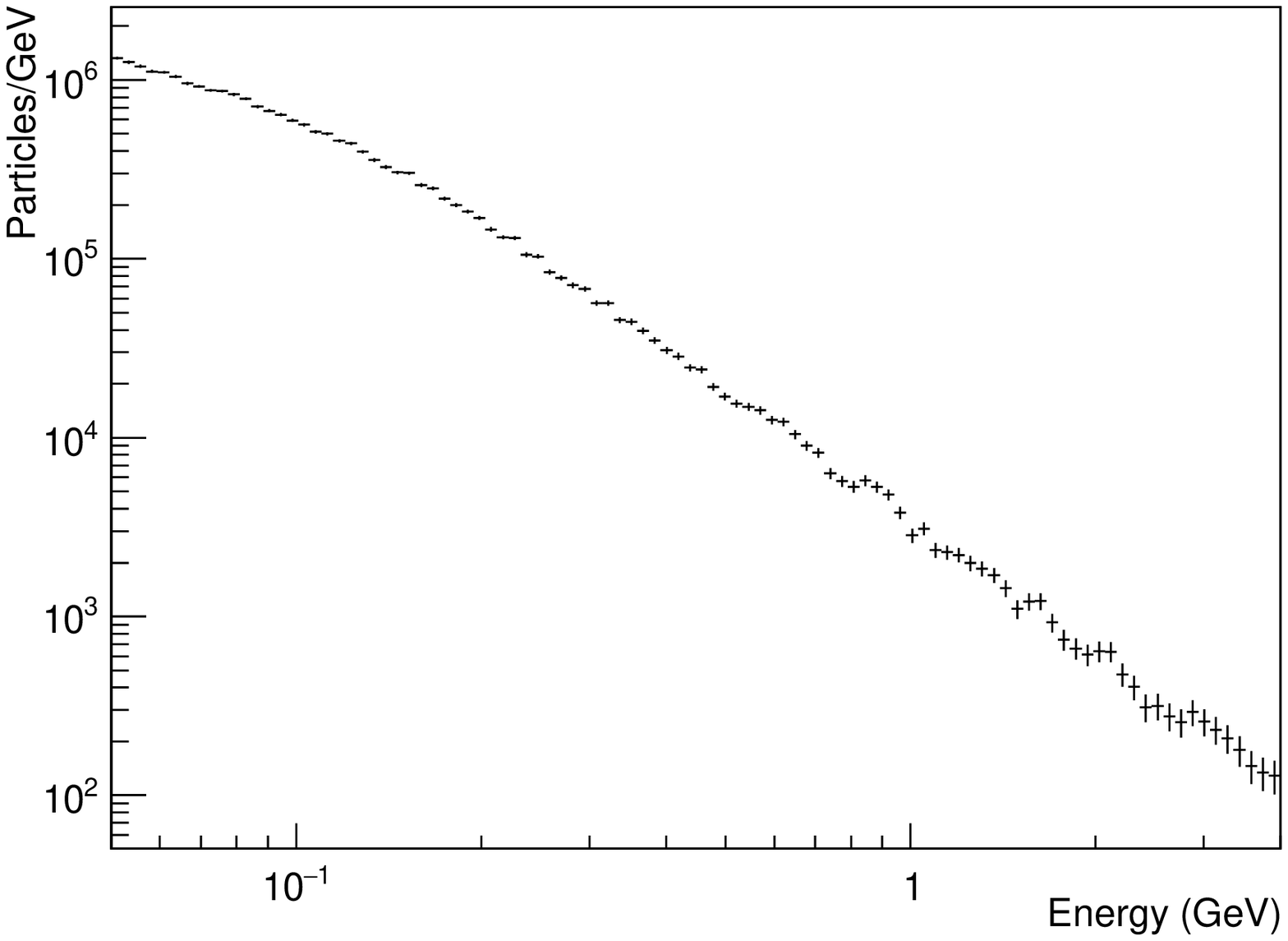}}
  \caption{(a) Trapped proton distribution as the function of kinetic energy and 
  distance of initial position of the trajectory. (b) Integrated flux distribution 
  of the trapped protons.}
  \label{fig:kedist}
\end{figure}

In Fig. \ref{fig:kedist}a, we show the trapped proton distribution with respect 
to the kinetic energy and the distance of its generation. The overall energy
spectrum is shown In Fig. \ref{fig:kedist}b. While the total 
entrapment efficiency of the CRAND protons in the considered energy range and 
altitude limit can be estimated from total number of simulated neutrons and 
entrapped protons given in Sec. \ref{ssec:ptrans} which is about 26\%, a 
relative calculation of the energy dependent entrapment efficiency was also 
done. This shows about 40\% lesser trapping efficiency at the lower energy end 
in comparison to that at the higher end and the efficiency increases almost 
monotonically towards the higher energy end.

\begin{figure}
  \centering
  \noindent\includegraphics[width=0.49\textwidth]{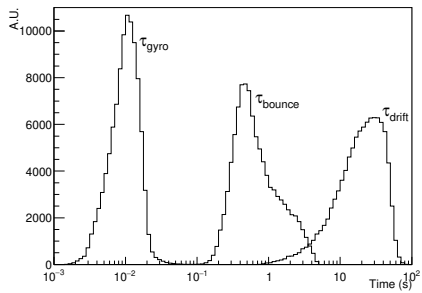}
  \caption{Distribution of temporal scales for three types of motions of all
  the simulated trapped protons under the effect of geomagnetic field.}
   \label{fig:timesc}
\end{figure}

We verified the trapped proton trajectories to fulfill the adiabatic conditions, 
particularly checking the hierarchy of the temporal scales:
\begin{equation}
\tau_{gyro} \ll \tau_{bounce} \ll \tau_{drift},
\end{equation}
where $\tau_{gyro}$, $\tau_{bounce}$ and $\tau_{drift}$ are time scales
associated to gyration, bounce and drift motions, respectively. $\tau_{gyro}$
($=\frac{2 \pi m_p}{q B}$; where $m_p$ proton mass, $q$ charge, $B$ local 
magnetic field) is calculated from the magnetic field value at the initial
position of the proton trajectory. While, $\tau_{bounce}$ and $\tau_{drift}$ 
are calculated by averaging the bounce periods and drift periods respectively, 
over the whole trajectory, obtained from the simulated trajectory information. 
Figure \ref{fig:timesc} shows the $\tau_{gyro}$, $\tau_{bounce}$ and 
$\tau_{drift}$ distribution for all the simulated trajectories.

\subsection{Adiabatic invariants}
\label{ssec:adinv}
Here, in this work, we have not considered the gradual energy loss, nuclear
scattering or radial diffusion of the trapped particles. So the adiabatic 
invariants of the trapped particles are conserved. The versions of the 
adiabatic invariants considered are as follows \citep{roed70}:
\begin{equation}
M = \frac{\gamma^2 m v_{\perp}^2}{2B},
\label{eq:M}
\end{equation}

\begin{equation}
K = \int_{s_m}^{s'_m} \left(B_m - B(s)\right)^{1/2} ds,
\label{eq:K}
\end{equation}

\begin{equation}
L = \frac{2 \pi \mu_E}{R_E \Phi}.
\label{eq:L}
\end{equation}

The first invariant $M$ is calculated using the particle velocity and magnetic
field value at the initial position of the track. The second invariant $K$ is
calculated by integrating the magnetic field along the guiding center path
between two successive mirror points. The average $K$ value over the whole
trajectory has been used as the second invariant of the trajectory. We
consider $L$ value as the third invariant. It is directly calculated from
distance of the trajectory from the Earth's center at the geomagnetic
equator, where the track is at the highest position. We subsequently calculate
$\Phi$, the magnetic flux inside the drift shell from Eq. \ref{eq:L}, where
$R_E$ is Earth's radius and $\mu_E$ Earth's magnetic dipole moment (arbitrarily
taken for 2008 A.D. to compare with the results from {\it PAMELA} measurements).

\begin{figure}
  \centering
  \noindent\subfloat[]{\includegraphics[width=0.33\textwidth]{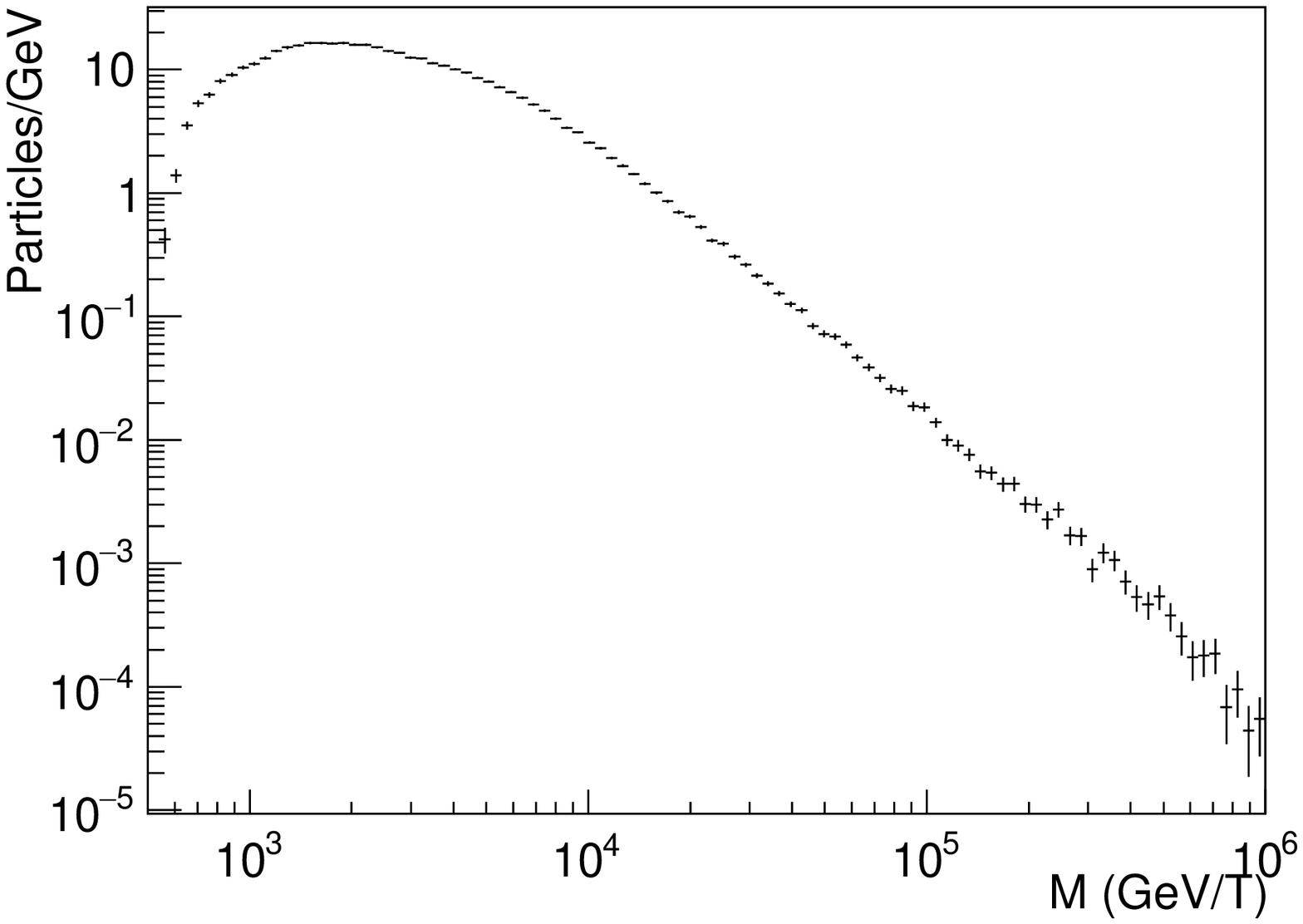}}
  \noindent\subfloat[]{\includegraphics[width=0.33\textwidth]{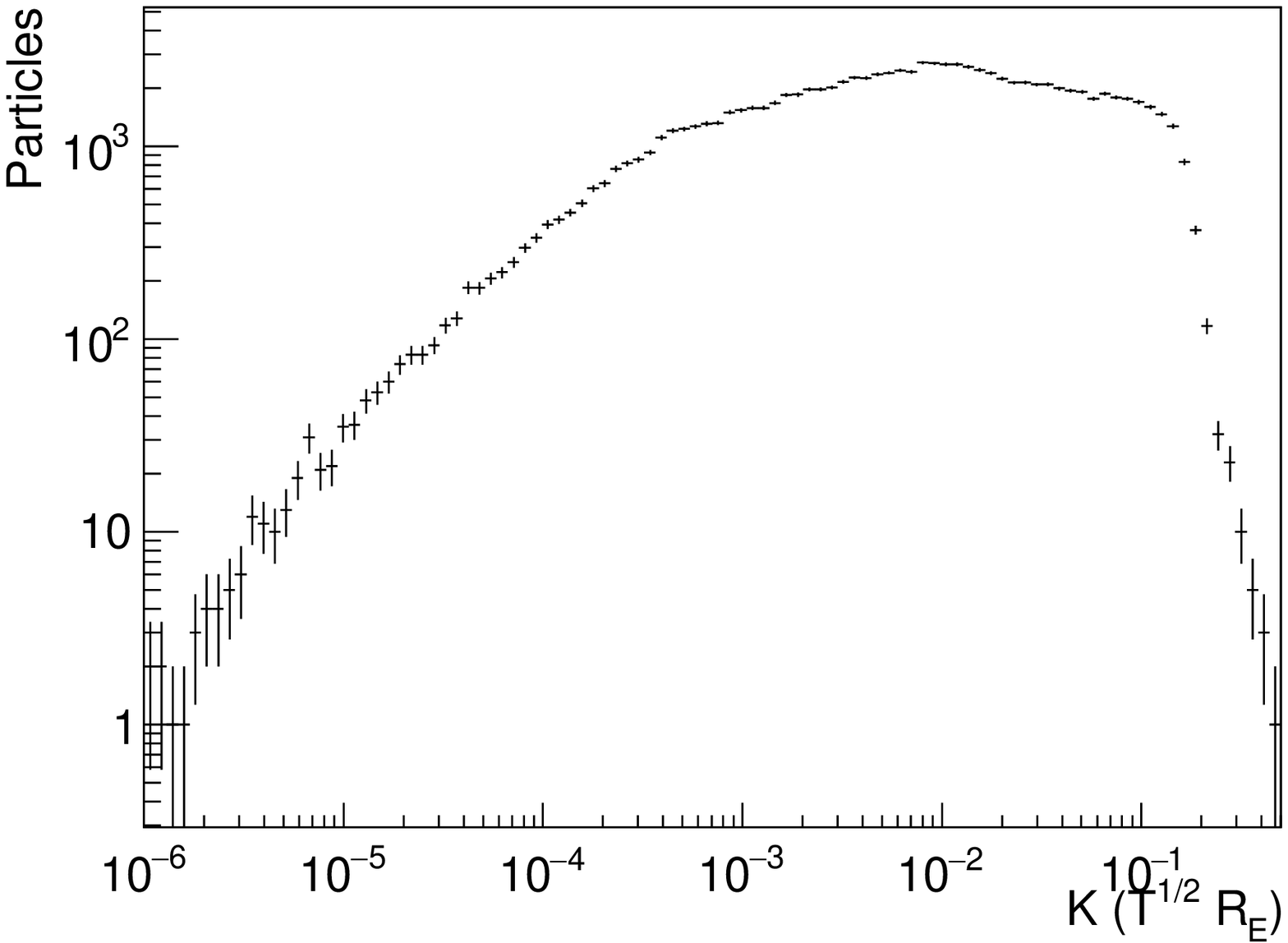}}
  \noindent\subfloat[]{\includegraphics[width=0.33\textwidth]{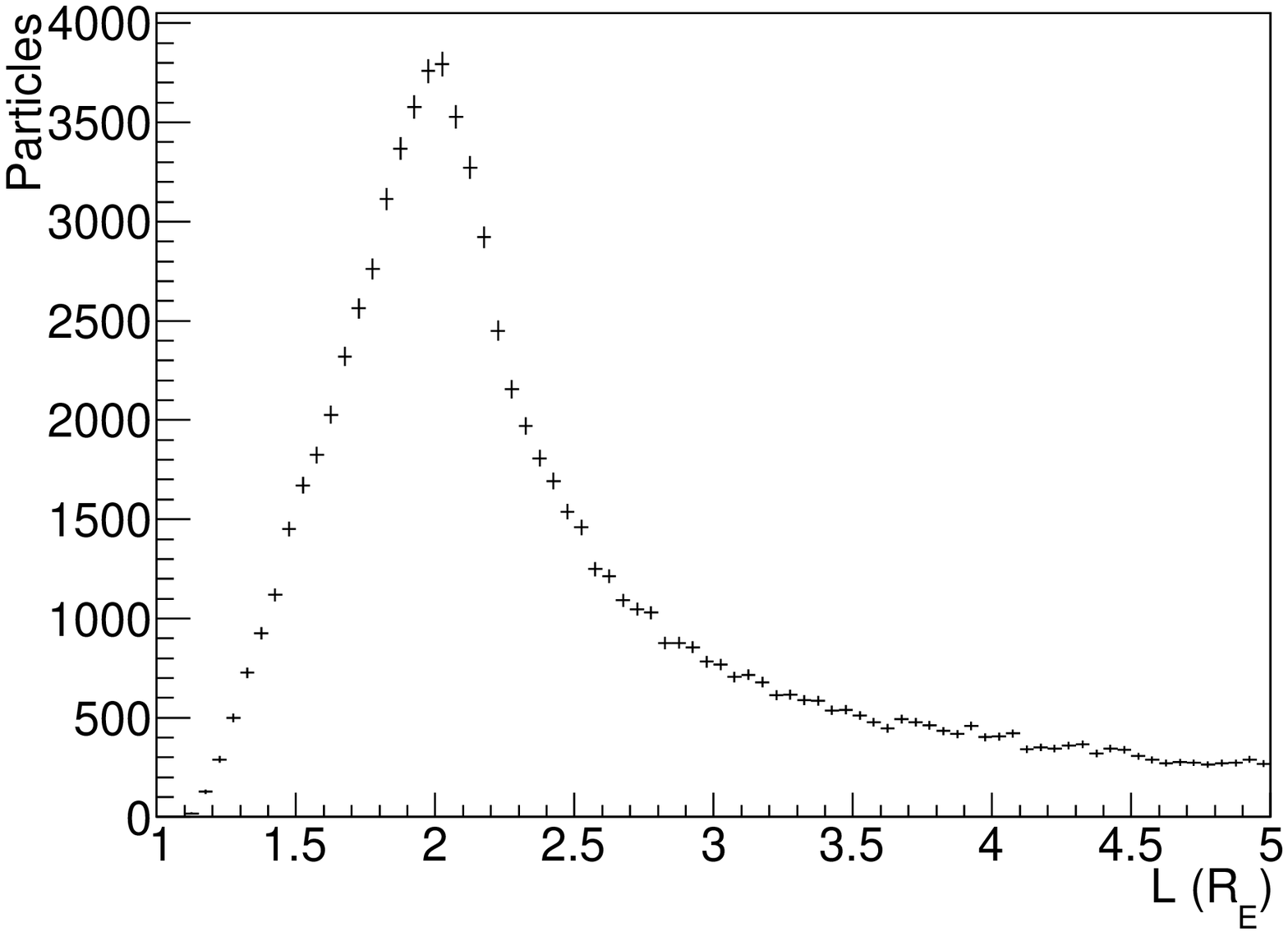}}
  \caption{Distribution of adiabatic invariants (a) M, (b) K and (c) L values
  for all the simulated trajectories.}
  \label{fig:adinv}
\end{figure}

The distribution of the three adiabatic invariants, for all the simulated particles 
are shown in Fig. \ref{fig:adinv}. The drop in population at extreme low $M$ 
values may be attributed to the scarcity of trapped particles at the lowest 
values of initial pitch angle as shown in Fig. \ref{fig:padist}b. The cut-offs 
apparent in the $K$ plot at around $10^{-1}$ $T^{1/2} R_E$ and in $L$ plot at 
2 $R_E$ are due to the limit used in the simulation for the particle generation 
distance. However, we calculated the trapped particle trajectories beyond this 
distance (up to the average magnetopause distance at 10 $R_E$).

\subsection{Spatial distribution of trapped protons}
\label{ssec:spdist}
\begin{figure}
  \centering
  \noindent\subfloat[]{\includegraphics[width=0.5\textwidth]{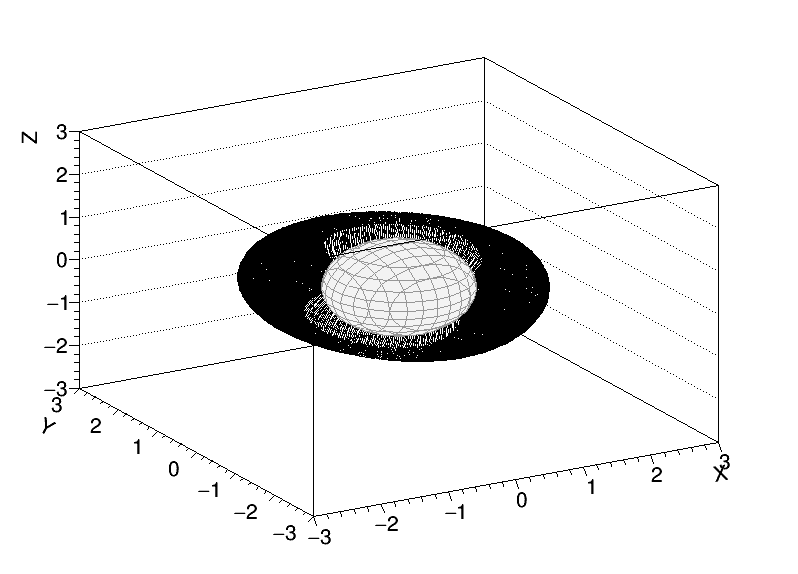}}\\
  \noindent\subfloat[]{\includegraphics[width=0.4\textwidth]{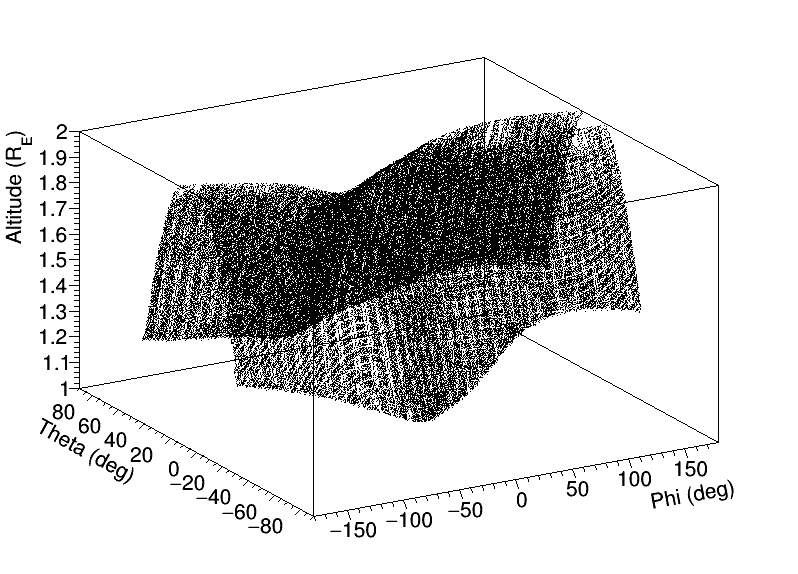}}
  \noindent\subfloat[]{\includegraphics[width=0.4\textwidth]{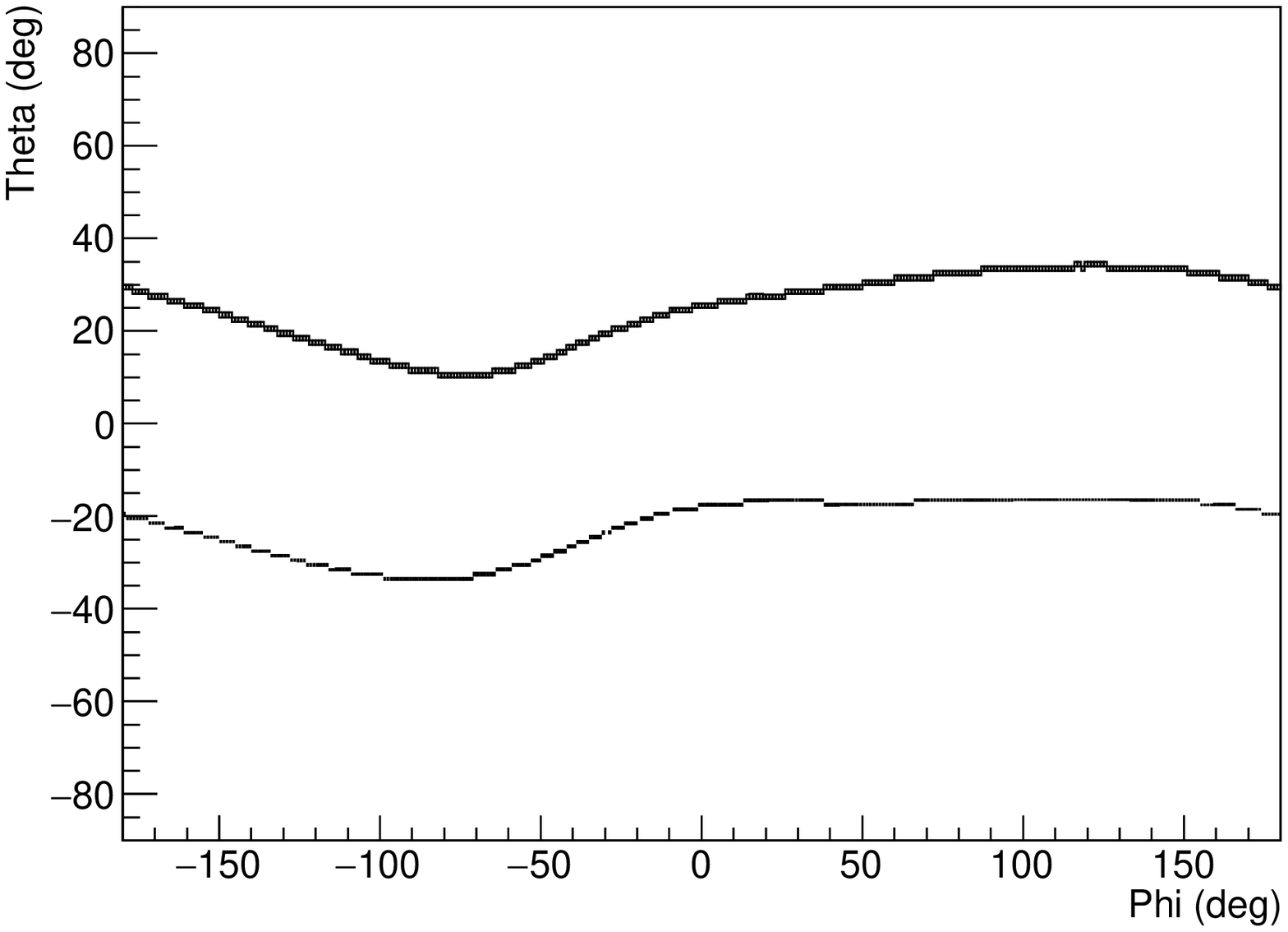}}
  \caption{(a) Example of a trapped particle track (guiding center) for a 288.7
  MeV proton started at 1.66 $R_E$ with initial pitch angle 135.9$^{\circ}$.
  (b) Same track represented in the grid view of geocentric coordinate system.
  (c) Distribution of the drift shell interception locations with a spherical
  surface at 1.7 $R_E$.}
  \label{fig:track}
\end{figure}

A stably trapped particle track creates a drift shell surrounding the Earth and
can be detected from any position in the continuous shell given a sufficiently 
long entrapment time. So, from the simulation point-of-view, we can spatially 
distribute one event to contribute all over the positions throughout the drift 
shell, improving the particle statistics without using extra computational 
resource for simulating more events. However, we must interpolate the 
trajectory to represent the drift shell. In Fig. \ref{fig:track}a, we show, 
for example, the guiding center trajectory of a 288.7 MeV proton trapped in the 
geomagnetic field started from a position at 1.66 $R_E$ altitude and with 
initial pitch angle 135.9$^{\circ}$. The trajectory chosen mere arbitrarily 
just to present the analysis procedure. For the analysis purpose, we replotted 
the trajectory in the geocentric coordinate system ($\phi$, $\theta$ and elevation 
from Earth's center), with $1^{\circ} \times 1^{\circ}$ grid in $\phi$/$\theta$ 
and dividing the elevation range of 1--2 $R_E$ in 320 bins (giving $\Delta Alt$ 
$\sim$ 20 km). The trajectory plot is shown in Fig. \ref{fig:track}b which 
represents the drift shell for the corresponding track. Then we can calculate 
the distribution of the drift shell interceptions at each layer of different 
elevation (measurement altitude) and interpolate the interception locations. 
This is shown in Fig. \ref{fig:track}c, for example at the measurement surface 
at 1.7 $R_E$. We adopted this method to ensure that, for the simulated 
trajectories, where the azimuthal gap between two subsequent bouncing tracks 
are larger (than the considered phi grid, i.e., 1$^{\circ}$ here), we don't 
have to rotate the trajectory around Earth for too many times, which can be 
computationally expensive. Instead, we fill the gaps (between subsequent 
locations of track interceptions) by interpolation as shown in Fig. 
\ref{fig:track}c.

\begin{figure}
  \centering
  \noindent\includegraphics[width=\textwidth]{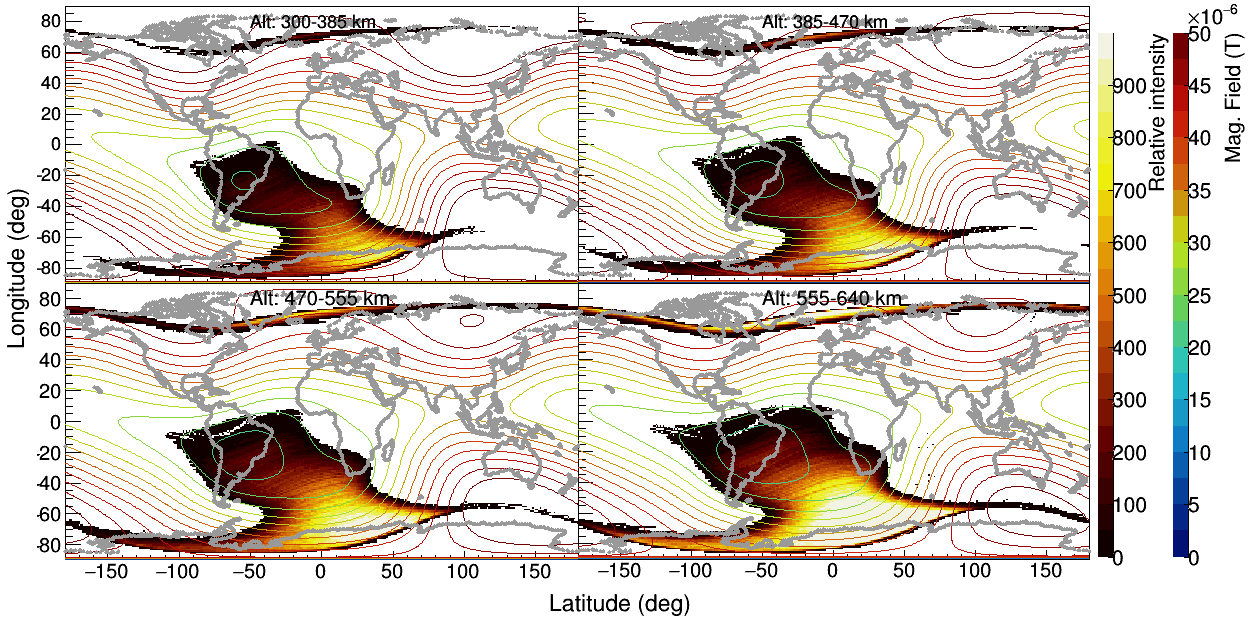}
  \caption{Distribution of trapped proton trajectories from CRAND in geographical
  latitude/longitude at four different altitude ranges. Geomagnetic field
  contours at the corresponding heights has been drawn overlaid on the proton
  distribution plots.}
   \label{fig:spdist}
\end{figure}

Then we calculate the overall contribution of all the simulated trapped 
trajectories at different measurement altitudes. In Fig. \ref{fig:spdist}, for 
example, we show the spatial distribution of proton trajectories integrated 
over four different altitude ranges at LEO: 300--385, 385--470, 470--555 and 
555--640 km. It is evident from these plots that the high energy trapped 
protons in the considered energy range and with all pitch angles are largely
concentrated at north-east of the geographic south pole below the African 
peninsula. 

\subsection{Flux normalization}
\label{ssec:norm}
The trapped proton intensity at some fixed location (on a satellite orbit) and
at a given time (of measurement) can be expressed as a function of kinetic 
energy ($E$) and time, in the following form \citep{sele01}:
\begin{equation}
j(E, t) = v \int_{t_0}^t S(E', t') \exp\left(-\int_{t'}^t
\left(\frac{\partial}{\partial E''} \frac{dE}{dt} + \frac{1}{\tau}\right)
dt'\right),
\label{eq:intensity}
\end{equation}
where $v$ is the speed of the particle corresponding to the kinetic energy, 
$S$ is the source rate, $\tau$ is the trapping lifetime and $\frac{dE}{dt}$ 
represents the continuous energy change of the particle. At the initial time 
$t$ = $t_0$, $j$ = 0, when $E$ corresponds to maximum energy of the trapped 
proton. However, in this work, we consider crude approximation of no gradual 
energy loss (through ionization; to free electrons or adiabatic energy change 
caused by secular geomagnetic variation) and no trapped particle loss through 
nuclear scattering or due to the effect from geomagnetic storms. So, 
$\frac{dE}{dt}$ = 0 and $\tau$ = $\infty$; we only have the time integration 
over the source term, which, in this present case, is the GCR contribution 
through CRAND mechanism. This approximation has severe implication over the 
absolute flux calculation of the trapped particles but reveals the relative 
flux distribution anyway.

The normalization factor of the simulation events to calculate the trapped
proton flux can be given by:
\begin{equation}
NF = \frac{\int_{T_s} \int_{A_s} \int_{\Omega_s} \int_{E_s} \frac{dN}{dE} dE
d\Omega da dt}{N_{sim} \int_{A_m} \int_{\Omega_m} d\Omega da},
\label{eq:norm}
\end{equation}
where $\frac{dN}{dE}$ is the source function of the albedo neutron energy
spectrum given by Eq. \ref{eq:neutspec} and $N_{sim}$ (= 10$^6$) is the total
number of simulated events giving trapped particles. Integration of the source
function is done over the energy range $E_s$ = [50 MeV, 4 GeV], over 2$\pi$
solid angle ($\Omega_s$) and over the area ($A_s$) of albedo neutron
generation surface (spherical surface at 100 km from Earth's surface).
Integration for the part at the denominator, corresponds to the measurement
of the trapped protons and is carried over 4$\pi$ solid angle ($\Omega_m$) and
over the spherical surface area $A_m$ at the measurement altitude ($\sim$ 500 km)
considering the average {\it PAMELA} orbit. The time limit $T_s$ , which
indicates the accumulation time of the trapped particles, is undefined here.
Since, we are not considering the loss of the trapped particles, the average
value of $T_s$ can only be inferred using the information of the measured flux.
In this calculation, considering a gross comparison with the {\it PAMELA} measurement
of the trapped proton flux, we get $T_s$ $\sim$ 0.1 year. This value is somewhat
lower in comparison to the trapped proton residence time range mentioned in
\citet{sele07} as less than a year to $\sim$ 4000 years. But this can be 
justified for the high energy particles (with smaller drift time scale) we are
considering here and since we are ignoring the particle loss effects.

\section{Comparison with PAMELA Measurements}
\label{sec:pamcom}
We compare the simulation results with the measurement of trapped proton
fluxes at LEO by the {\it PAMELA} experiment \citep{adri15}. {\it PAMELA}
measured the trapped proton fluxes at $350\div610$ km altitude between July,
2006 and September 2009. The proton trajectories were reconstructed using
tracing programs based on numerical integration method \citep{smar00, smar05} 
to select the stably trapped trajectories used for the analysis. A total of 
7.3 $\times$ 10$^6$ stably trapped particles were sampled for the analysis out 
of total 9 $\times$ 10$^6$ detected events (including quasi- and un-trapped 
trajectories). We considered similar order of event number in the current 
simulation to get comparable statistics in the results.

\begin{figure}
  \centering
  \noindent\includegraphics[width=\textwidth]{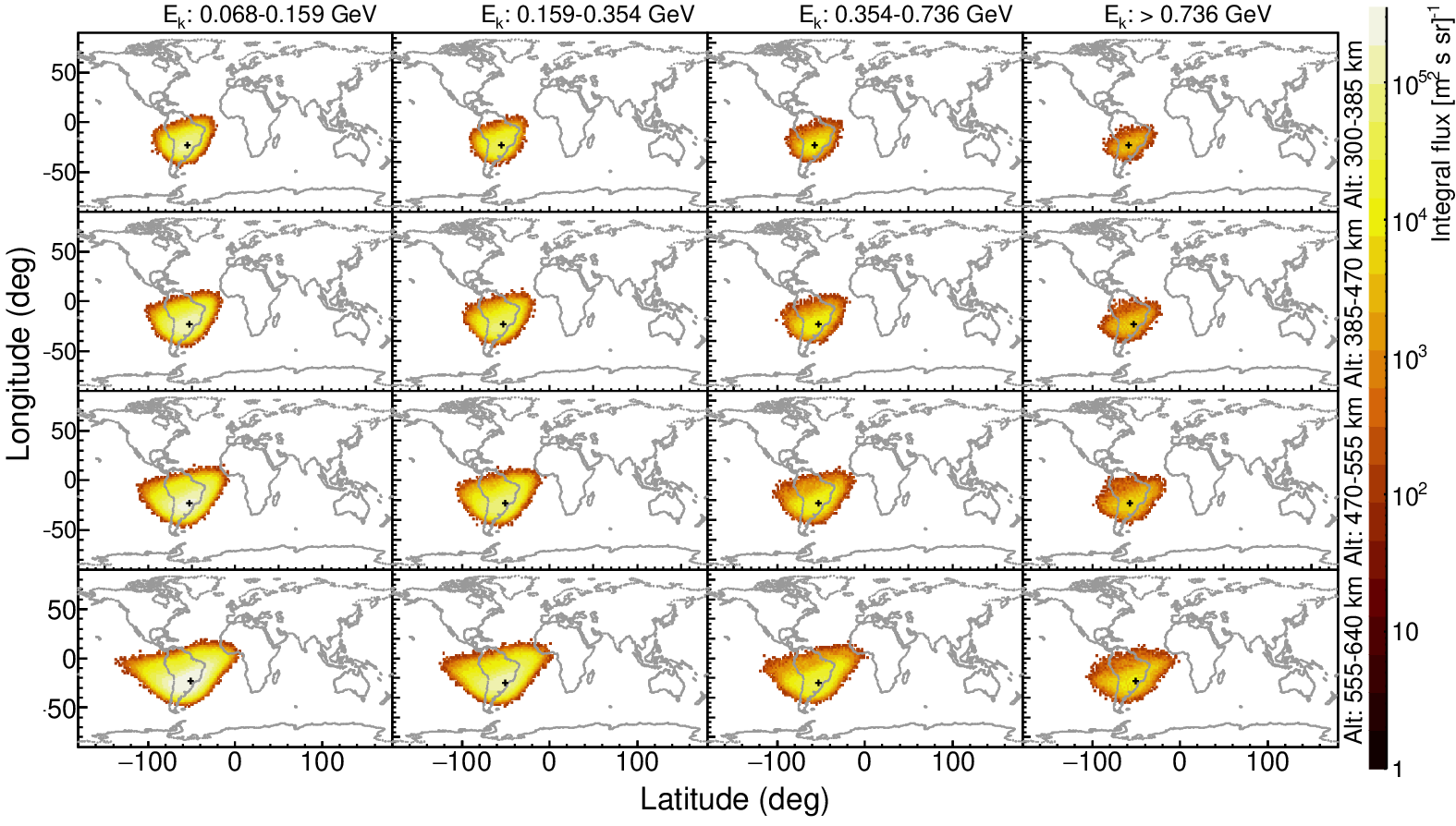}
  \caption{Simulated trapped proton integral flux distribution, integrated over 
  the pitch angle range covered by {\it PAMELA}, in geographical coordinates, 
  at different altitudes (rows) and energy ranges (columns). The peak of the 
  distributions are indicated by the plus signs.}
   \label{fig:spdistpam}
\end{figure}

In Fig. \ref{fig:spdistpam}, we show the distribution of trapped proton flux
integrated over the pitch-angle range covered by {\it PAMELA}, at different
altitudes and energy ranges. We refer to the Fig. 1 of \citet{adri15} for the
comparison of the distribution. We rebinned the $\theta/\phi$ distribution in
$2^{\circ} \times 2^{\circ}$ grid for the comparison. We also added an
uncertainty of $\sigma = 3^{\circ}$ to the simulated track positions (in 
$\theta/\phi$), to account the uncertainty in payload orientation measurement
and angular resolution of the detector. We sampled the detector
location over all the region at different altitude (but limited by the 
satellite orbit inclination of 70$^{\circ}$) and obtained the trajectories 
within the acceptance angle of the detector, by calculating the angle between 
the detector direction (towards the local zenith) and trajectory direction 
(obtained from the local pitch angle of the trajectory) at the location of the 
detector. However, this approach averages the calculation over the gyro-phase 
angle and neglects the east-west effect due to atmospheric absorption at 
larger gyro-radius. The simulated 
result of the trapped proton distribution map precisely reproduces the measured 
distribution by {\it PAMELA} which is concentrated around the South Atlantic 
Anomaly (SAA) region. Relation of the area coverage with the altitude and 
energy range is also apparent from the plot. In particular, the feature 
reported in \citet{adri15}, that the proton flux is more concentrated in the 
southeast part of SAA is also apparent here. To make this feature more apparent 
we also indicate the peak points of the distribution on the same plot. The 
explanation of the peaks shifted towards the southeast can be visible from Fig. 
\ref{fig:spdist}, which shows that the trapped protons, irrespective of 
{\it PAMELA} acceptance angle, are concentrated near the south-polar region 
below the African peninsula. We have already mentioned in Sec. \ref{ssec:norm}, 
the absolute value of the integral flux depends on the accumulation time of 
the trapped particles, which require to include the loss effects to calculate 
it. Since, in this work we did not consider the loss effects, to present the 
integral flux we simply scaled the value by varying the $T_s$ value to grossly 
match the {\it PAMELA} measurement.

\begin{figure}
  \centering
  \noindent\subfloat[]{\includegraphics[width=0.5\textwidth]{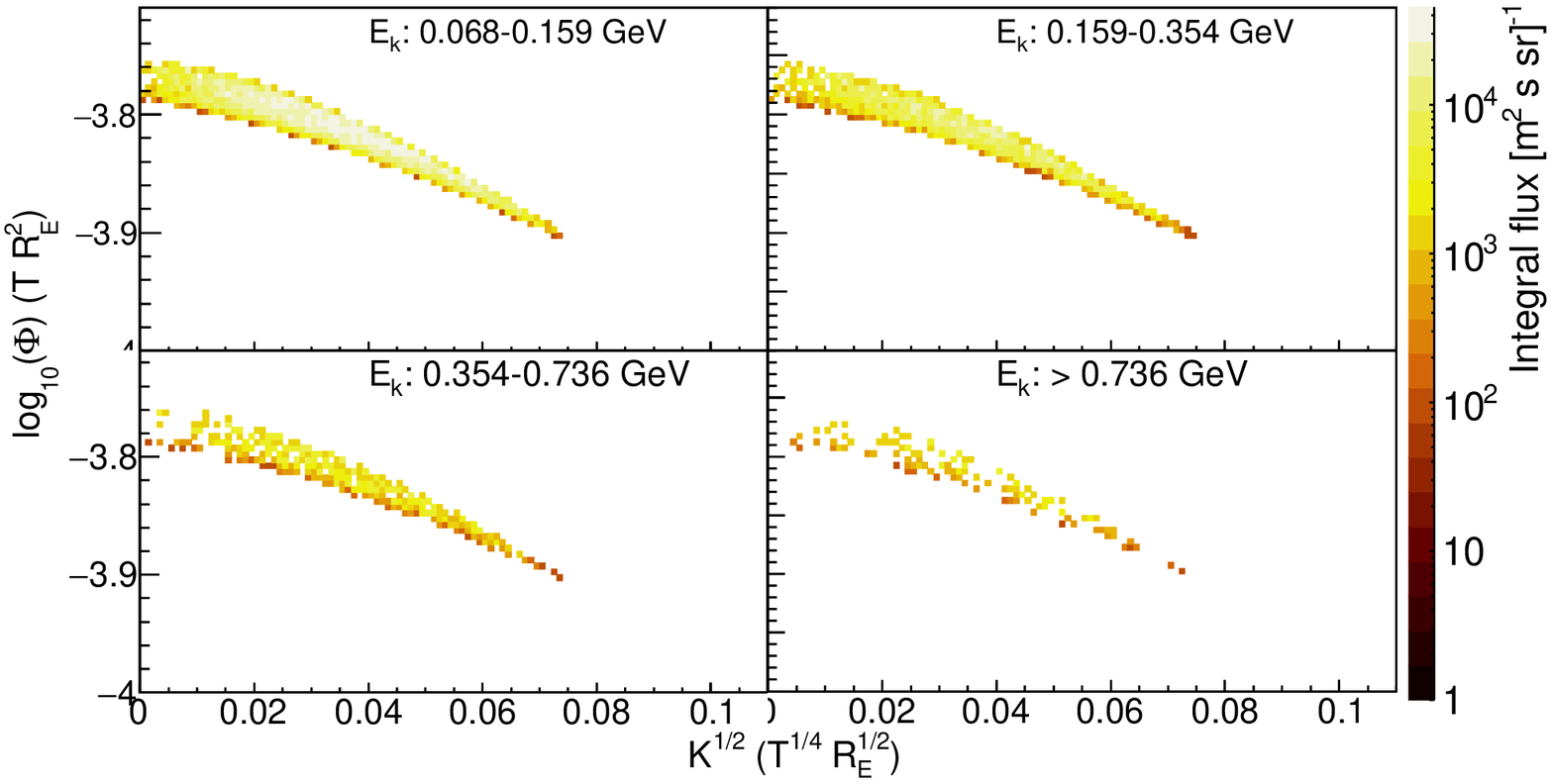}}
  \noindent\subfloat[]{\includegraphics[width=0.5\textwidth]{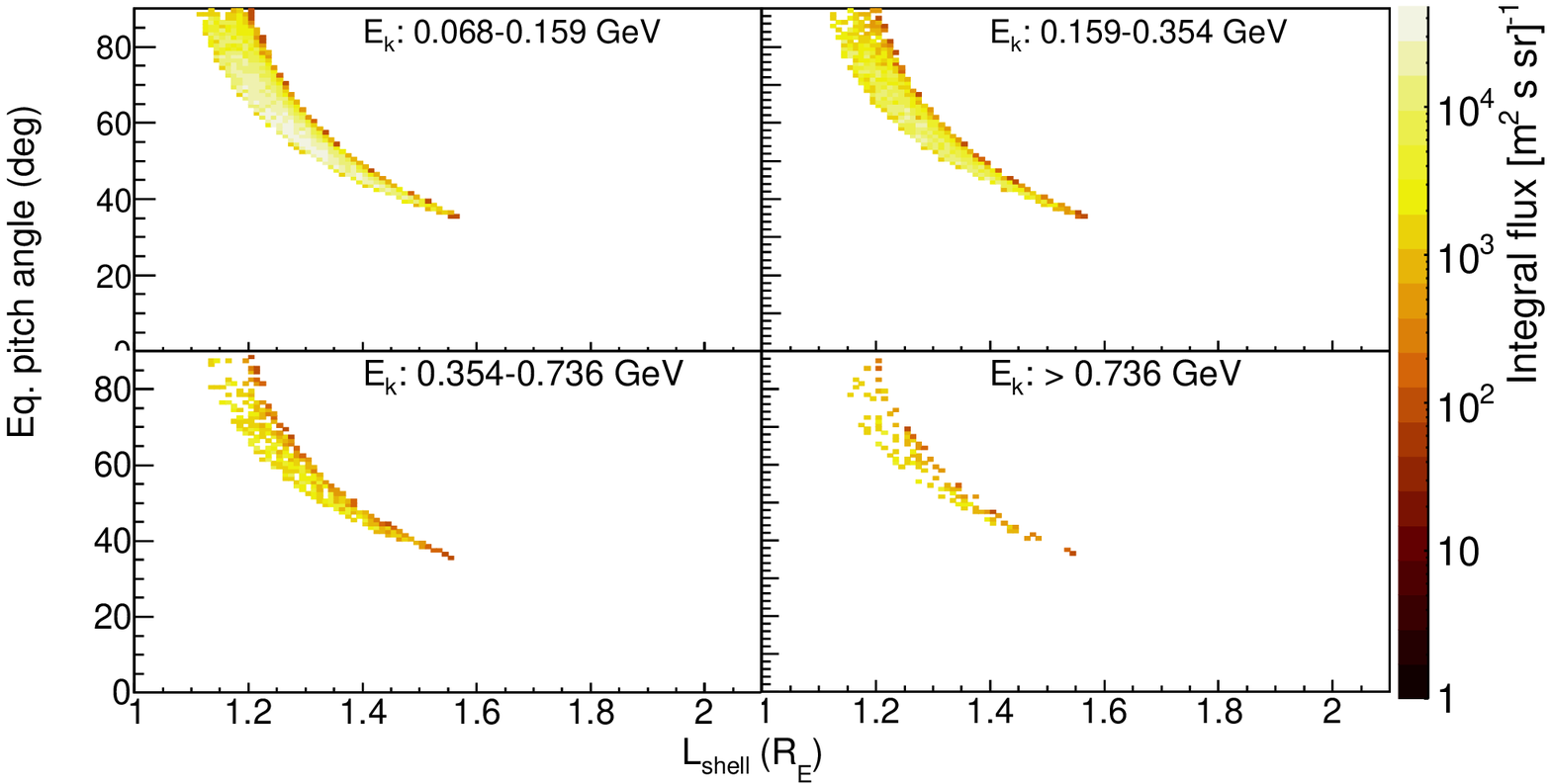}}
  \caption{(a) Proton integral fluxes as a function of K and $\Phi$ at different
  kinetic energy bins in the pitch angle range covered by {\it PAMELA}. (b) The 
  similar distribution as a function of L-shell parameter and equatorial pitch angle.}
   \label{fig:kphipam}
\end{figure}

For additional comparison with the measured data, we also plot the integral 
proton flux with other parameters than spatial distribution. Figure 
\ref{fig:kphipam}a shows the proton integral fluxes as a function of second 
and third adiabatic invariants ($K$ and $\Phi$), for the particles in 
{\it PAMELA} field of view, at different kinetic energy ranges. We plotted the 
result in $K^{1/2}$ and $\log_{10}\Phi$ coordinates for better resolution (and 
for the sake of direct comparison with {\it PAMELA} results). Figure 
\ref{fig:kphipam}b shows the same fluxes as a function of equatorial pitch 
angle and $L$ value. These results can be compared with the results from the 
{\it PAMELA} measurements shown in Fig. 2 and 3 of \citet{adri15}. While the 
overall distribution agrees well with the observed data, the comparisons show 
some differences in the extent of the energy dependent data patches in $K$ and 
$\alpha_0$. These differences can be attributed, as discussed in 
\citet{adri15}, to the spacecraft orbit which constraint the observation of 
the equatorial mirroring protons only for L values up to $\sim$ 1.18 R$_E$. 
However, one thing should be mentioned regarding the simulation aspect --- 
here in this work we optimized the algorithm to gain statistically in the spatial 
distribution of the trapped particles where one trajectory can contribute to 
the entire drift shell region. But, since this optimization has no effect on 
increasing the number of trajectories, the analysis concerning the parameters 
attributed to a trajectory has little gain from the algorithm. We need to simulate 
more particles for a better comparison of these aspects shown in Fig. 
\ref{fig:kphipam}.

\section{Conclusions}
\label{sec:conc}
We conducted a thorough simulation to calculate the trapped proton radiation
contribution at LEO, through the CRAND mechanism. This is the principal
contributor to the proton radiation at this region for proton energy $\gtrsim$
50 MeV. For this purpose we developed a simulation framework to calculate the
secondary particles from the atmospheric interaction of the GCRs. The protons,
produced from the $\beta$-decay of the albedo neutrons from these GCR
interactions, are then geomagnetically trapped and their trajectories have been
calculated using the guiding center equation of the particles. We studied the
spatial as well as the phase-space distribution of the trapped protons and
compared the results with the observation by {\it PAMELA} mission, giving
satisfactory agreement. This study also helps to understand certain features of
the measured flux distribution. However, in this calculation we considered only
the source term, neglecting the loss of trapped particles over time, due to
various scattering and energy loss processes. Consideration of the loss term is
important for the complete understanding of the radiation distribution and will
be added in the future extension of the work. 

While, in this work we have particularly concentrated on the trapped proton 
distribution at LEO and to compare the results with {\it PAMELA} measurement, 
the same simulation framework can be used, to study the overall inner radiation
belt by investigating the spatial distribution, trapping limits etc. We can 
also consider the contributions from other sources to the radiation belt like 
solar proton events with some modification to the source particle generation
and the results can be compared to the recent empirical models like AP9 
\citep{gine13} and that given by \citet{sele07}.

The simulated result agrees well with the observation which is evident from
Fig. \ref{fig:spdistpam}. This flux distribution is mainly due to the trapped 
trajectories with L values less than 1.6. However, a closer inspection of the 
overall flux distribution shown in Fig. \ref{fig:spdist} reveals that there are 
two different populations of trajectories that comprise the distribution. The 
stably trapped particles with L $<$ 2 are visible near the SAA region. 
Trajectories in the other part comprising the horns and deposition near the 
south pole mainly show higher L values whose entrapment are theoretically 
perplexed for such high energy particles. The origin and nature of these 
``trapped'' trajectories could be better revealed by a more detailed 
investigation of the morphology of these trajectories which has not been done 
yet in this work and is pending for the future extension of this work.


\section{Acknowledgments}
We express our sincere gratitude to M. Boezio (INFN, Trieste) and 
A. Bruno (NASA/GSFC) for the helpful discussions. This work has been done 
under partial financial support from the Science and Engineering Research 
Board (SERB, Department of Science and Technology, Government of India) 
project no. EMR/2016/003870. We also thank the Higher Education department 
of West Bengal, for a Grant-In-Aid which allowed us to carry out the research 
activities at ICSP. 

\bibliographystyle{model5-names}
\biboptions{authoryear}
\bibliography{reference}

\end{document}